\documentstyle[12pt]{article}
\newcommand{\pr}{\paragraph{}}
\newcommand{\be}{\begin{equation}}
\newcommand{\ee}{\end{equation}}
\newcommand{\bea}{\begin{eqnarray}}
\newcommand{\nn}{\nonumber}
\newcommand{\eea}{\end{eqnarray}}

\newcommand{\nk}{\noindent}
\begin{document}

\begin{titlepage}
\begin{flushright}
CERN-TH.7000/93 \\
CTP-TAMU-66/93 \\
ACT-23/93 \\
ENSLAPP-A-445/93 \\
OUTP-93-26P \\
\end{flushright}
\begin{centering}
\vspace{.1in}
{\large {\bf A Liouville String Approach to Microscopic
Time and Cosmology}} \\
\vspace{.2in}
{\bf John Ellis$^a$},
{\bf N.E. Mavromatos$^b$},
{\bf D.V. Nanopoulos$^{c,\diamond}$}\\
\vspace{.03in}
\vspace{.1in}
{\bf Abstract} \\
\vspace{.05in}
\end{centering}
{\small
    In the non-critical string framework that we have proposed recently,
the time $t$ is identified with a dynamical local renormalization group
scale, the
Liouville mode,
and behaves as a statistical evolution parameter,
flowing irreversibly from an infrared fixed
point - which we conjecture to be a topological string phase - to an
ultraviolet one - which corresponds to a static critical string vacuum.
When applied to a toy
two-dimensional
model of space-time singularities, this formalism yields an
apparent renormalization
of the velocity of light, and a $t$-dependent form
of the uncertainty
relation for position and momentum of a test string. We speculate
within this framework on
a stringy alternative to conventional
field-theoretical
inflation, and the decay towards zero of the cosmological
constant in a maximally-symmetric space. }

\begin{flushleft}
CERN-TH.7000/93 \\
CTP-TAMU-66/93 \\
ACT-23/93 \\
ENSLAPP-A-445/93 \\
OUTP-93-26P \\
\end{flushleft}
\vspace{0.2in}
%
%
$^a$ Theory Division, CERN, CH-1211, Geneva, Switzerland,  \\
$^b$ Theoretical Physics Laboratory,
ENSLAPP,
Chemin de Bellevue,
Annecy-le-Vieux, France, and  \\
S.E.R.C. Advanced Fellow, University of Oxford, Dept. of Physics
(Theoretical Physics),
1 Keble Road, Oxford OX1 3NP, United Kingdom,   \\
$^c$ Center for Theoretical Physics, Dept. of Physics, \\
Texas A \& M University, College Station, TX 77843-4242, USA,
and \\
Astroparticle Physics Group,
Houston Advanced Research Center (HARC),
The Mitchell Campus, The Woodlands, TX 77381, USA\\
\vspace{0.01in}
$^{\diamond}$ {\bf Opening lecture} at {\it the HARC
Workshop on Recent Advances in the
Superworld}, The Woodlands, Texas, April 14-16, 1993


%
\vspace{0.01in}
\begin{flushleft}
November 1993 \\
\end{flushleft}

\end{titlepage}
\newpage
\section{Introduction}
\pr
The subject of
fundamental constants  in field or string theory
is a fascinating challenge for any physicist. Unfortunately,
despite vigorous attempts, the topic
remains controversial today.
We would not exaggerate if we said that it is
still at a very primitive stage of understanding.
Following Weinberg \cite{weinb},
we can
define
as ``fundamental constants'' a list of constants whose ``value
we cannot calculate in terms of more fundamental constants \dots
because we do not know of anything more fundamental \dots" .
Thus,
Weinberg concludes that  ``the membership of
such a list
reflects our present
understanding of fundamental physics."
\pr
We shall not attempt in this talk to offer any solutions to
this fundamental problem in field theory. Instead,
we shall examine
the consequences of a recently-proposed
approach \cite{emnqm} to the origin of
time in non-critical String Theory \cite{aben3,aben} for some
``constants'' in string theory, namely the speed of light $c$,
and the fundamental (minimal) length $\lambda _s$ derived from the
uncertainty principle \cite{venez}.
\pr
According to Veneziano \cite{ericeven}, string theory appears to have
no fundamental constants {\it
per se}, but these constants arise as a
result of spontaneous breaking of symmetries in the selection of
a ground state,
thereby leading a (string) physicist to conclude that
\cite{ericeven} ``the constants of the vacuum are the constants
of the world''.
To prove or disprove
this
statement
is a formidable
task.
Presumably,
to understand, in String Theory,
the origin
of the
fundamental constants that characterize the vacuum would
require first an answer to the important and still unresolved
issue of
what
(if anything) lifts
the huge degeneracy that
appears at present
to characterize
the string ground states.
This question is tied up with the question on the origin of
space-time.
\pr
As a
step in
this direction,
we have
formulated \cite{emnqm,emndollar}
time in string theory as a renormalization group scale on the
world-sheet, which we identify with the Liouville mode.
The connection of time with the Liouville field in non-critical
string theory has been considered in \cite{aben3,aben}
in order to incorporate an expanding Universe
in string theory.
In \cite{emnqm,emndollar}
we took one step further, and interpreted the Liouville
mode as a local renormalization group scale, which flows
irreversibly according to a generalization of
Zamolodchikov C-theorem \cite{zam}.
This scale
is
introduced
in theories with target space-time singularities  or, more generally,
event or
particle horizons,
as a result of the
breaking of conformal invariance by the
truncation of the
matter theory to observable local low-energy modes. In the toy
example of a two-dimensional black hole that we used to
develop
these
ideas \cite{emnnew,emndollar},
these low-energy propagating modes
are coupled to quasi-topological delocalized higher-level string modes,
as a result of infinite level-mixing gauge stringy
symmetries \cite{emn1}.
These discrete modes are believed to be remnants
of a topological phase in string theory \cite{emntop}.
The information
carried by these modes, which is hidden
from a low-energy observer who `measures'
using local scattering methods, is the
root of
the flow of time  and the non-equilibrium nature of the
low-energy field theory derived from the string.
As we have noted \cite{emnqm}, the fact that information
is lost to these hidden modes causes apparent
decoherence of the observable modes, reflected in a
collapse of their (part of the full) wave function that is
rapid for more complex systems.
\pr
In this scenario, the resulting low-energy two-dimensional
string-inspired theory
appears not to have any `fundamental' constants, however there are
dimensionful
parameters that relax to their {\it finite}
critical-string values
asymptotically as the Liouville time  $t \rightarrow \infty $. These
quantities
cannot be calculated in terms of anything more fundamental
in the model.
According to Weinberg's definition, then, the asymptotic
values of these quantities, which in this two-dimensional example
include the speed of light and the minimal uncertainty in
length, or $\hbar$,
would characterize the vacuum state and hence
would be the dimensionful
fundamental
constants of that ground state.
In addition,
there appears
in the vacuum
an
arbitrary expectation value of the dilaton field, which
could be considered as an arbitrary dimensionless constant,
the analogue of the fine structure constant in four-dimensional
models. In the non-equilibrium model the dilaton field
is a time-dependent scale factor that relaxes to its asympotic
(vacuum)
value.
We find it interesting,
but also quite natural,
that the non-equilibrium nature
of the non-critical Liouville string  induces
the appearance of constant quantities only
in the asymptotic
vacuum state via a
(quantum) relaxation process. After all, the best way of understanding
an equilibrium state in statistical mechanics is the study of
the approach to it by perturbing the relevant system away
from equilibrium.
This is
precisely the way we studied time in string theory. Using the
concept of `measurement' through localized fields, we have
perturbed the conformal string theory by operators that spoil
conformal invariance and, thus, introduced the concept of time as
an evolution parameter of the relaxation
process\cite{emnqm,emndollar,emnnew}.
\pr
As a byproduct  of
our approach we
speculate on
cosmological applications,
with emphasis on the possibility for a string alternative
to conventional field-theoretical
inflation. We also
deal with
the issue of the target-space cosmological
constant in generic $D$-dimensional maximally-symmetric spaces,
in models with a dilaton such as
certain no-scale supergravities \cite{noscale},
and
find \cite{emnnew} that it relaxes to zero
asymptotically.
\section{String-Inspired
Density Matrix Mechanics}
\pr
In this section we
review briefly the
formalism of first-quantized non-critical
strings, and its connection with the
Density Matrix Mechanics that we proposed
recently \cite{emnqm}. We emphasize
the
restoration of
criticality by turning a $(1,1)$
deformation
into an exactly marginal one by Liouville dressing, thereby
leading, in our interpretation,
to time-dependent backgrounds.
Consider a two-dimensional
critical conformal field theory model
described by an action $S_0(r)$ on the world sheet,
where the
$\{ r \}$ are matter fields
spanning a $D$-dimensional target manifold  of Euclidean
signature, that we
term
``space''.  Consider, now, a
deformation
\be
    S=S_0 (r) + g \int d^2 z V_g (r)
\label{nonmarg}
\ee
Here $V_g$
 is a (1,1) operator,  i.e. its anomalous dimension vanishes, but
it is not {\it exactly marginal}
in the sense that the operator-product
expansion coefficients $C_{ggg}$
of $V_g$ with itself are non-zero in any renormalization
scheme,
and hence
`universal' in the Wilsonian sense.
The scaling dimension $\alpha _g$
of $V_g$
in the deformed theory (\ref{nonmarg}) is, to $O(g)  $ \cite{ginsp},
\be
           \alpha _g =-g C_{ggg} + \dots
\label{scaling}
\ee
Liouville theory \cite{DDK}
requires  that scale invariance
of the theory (\ref{nonmarg}) be preserved. The non-zero
scaling dimension
(\ref{scaling}) would jeopardize
this, but
scale invariance
is
restored if
one dresses $V_g$ gravitationally on the world-sheet as
\be
\int d^2z g V_g (r)
\rightarrow
 \int d^2z g e^{\alpha _g \phi}V_g (r)
 =\int d^2z g V_g (r)
 -\int d^2z g^2 C_{ggg}V_g (r)\phi  + \dots
\label{dressed}
\ee
where $\phi$ is the Liouville field. The latter acquires dynamics
through integration over world-sheet covariant metrics,
$\gamma _{\alpha\beta}$,
after
conformal gauge fixing $\gamma _{\alpha\beta}=e^\phi {\hat
\gamma_{\alpha\beta}}$
in the way discussed in \cite{DDK,mm}.
Scale invariance is guaranteed through the definition
of renormalized couplings $g_R$ , given in terms of $g$ through
the relation
\be
    g_R\equiv g - C_{ggg}\phi g^2  + \dots
\label{renorm}
\ee
This
equation leads to the correct $\beta$-functions for $g_R$
\be
      \beta _g =-C_{ggg}g_R^2  + \dots
\label{beta}
\ee
implying a renormalization-group scale $\phi$-dependence of
$g_R$. The reader
might have noticed that above
we viewed
the Liouville field as a local scale on the world-sheet
\cite{emnqm,emndollar}. Local world-sheet scales
have
been
considered in the past
\cite{shore,osborn}, but the crucial difference of our Liouville
approach is that
this scale is made
dynamical  by being
integrated out
in the
path-integral. In this way, in Liouville strings the local dynamical
scale acquires
the interpretation of an additional target
coordinate. If the central charge of the matter theory is $c_m > 25$,
the signature of the kinetic term of the Liouville  coordinate
is opposite to that of the matter fields $r$, and thus the Liouville
field is interpreted as Minkowski target time \cite{aben3,aben}.
In this interpretation the renormalization group evolution of the
matter system alone, ignoring Liouville dynamics, is interpreted as
time evolution, in which the time appears as an `external' parameter.
 As argued in ref. \cite{kogan,emndollar}, the
actual time flow is
{\it opposite} to the usual renormalization
group flow, the latter being from ultraviolet to infrared
fixed points. This has to do with metastability interpretation
of the renormalization group flow in Liouville
strings \cite{kogan,emndollar}, where
the infrared fixed point is viewed as a `bounce' point of the flow
\cite{coleman}.
\pr
It is instructive
to review the properties of such a flow.
\pr
(i) Due to the assumed unitarity of the matter system, the flow
is {\it irreversible}, as can be seen using the
Zamolodchikov
$C$-theorem \cite{zam}.
The effective central charge (Zamolodchikov C-function \cite{zam})
$C(g)$ varies with time as
\be
   \partial _t C(g)=\beta ^i <V_i V_j > \beta ^j \equiv
\beta ^i G_{ij} \beta ^j \ge 0
\label{cfunct}
\ee
where $G_{ij}$ plays the r\^ole of a metric in coupling constant
space.
This implies a time {\it arrow} in all such models.

\pr
(ii) Defining a density matrix for the system, in a coupling
constant phase space $g^i$, $p_i$,
one has the following modified Liouville equation \cite{emnqm,ehns}
\be
\partial _t \rho =\{H,\rho \} + \beta ^i G_{ij} \frac{\partial \rho}
{\partial p_j}
\label{liouv}
\ee
where $\{ , \}$ denote the
appropriate Poisson bracket and $H$ is
a Hamiltonian function. The latter is defined in terms of the
effective action, which in string theories coincides with the
Zamolodchikov $C$-function \cite{mm2}
\be
     C(g)=\int dt p_i \beta ^i - H
\label{effeh}
\ee
The total probability $P=\int dg^idp_i tr \rho (g,p,t)$ is
conserved, since its rate of change is a boundary term
in phase space, and
the latter is
assumed to have no
boundary.
\pr
(iii) Quantization of the coupling constants $g^i$
is achieved
as follows: the
$g^i$ actually represent target-space  background
fields in string theory, and as such they are quantum field operators
in a string field theory concept, where summation over genera is
assumed. Hence
the situation is
similar to a two-dimensional
wormhole calculus.
{}From a quantum point of view, the Poisson brackets
$\{ , \}$ are then replaced by
appropriate commutators, and the couplings $g^i$ are viewed as Heisenberg
operators \cite{emnqm,emndollar}.
\pr
(iv) Entropy $S$ is defined in terms of the density matrix
$\rho$ as usual, by
\be
     S=-Tr \rho ln \rho
\label{entropia}
\ee
and it increases with time $t$
\be
           \partial _t S = \beta^i G_{ij} \beta ^j S \ge 0
\label{entrincre}
\ee
thereby implying, if any $\beta ^i \ne 0$,
non-equilibrium
for the non-critical
matter theory. The latter approaches
asymptotically
its critical state as an ultraviolet fixed point of the flow.
\pr
(v) Energy $H$ is conserved on the average, in this quantum version,
as a result of the renormalizability of the $\sigma$-model \cite{emnqm}
\be
   \partial _t << H >> \equiv
    \partial _t Tr H\rho = Tr (\rho \partial _t H )
    + Tr( H\partial _t \rho)=
    \partial _t (p_i\beta^i) =0
\label{energy}
 \ee
where we used (\ref{liouv}) and (\ref{effeh}). The vanishing
result is a consequence of the fact that
neither $p_i=\frac{\delta}{\delta g^i}$ nor $\beta ^i$
have an exlicit scale dependence in renormalizable theories.
However, fluctuations in the
energy are time-dependent. For instance,
\be
 \partial _t << H^2 >>= <<\frac{d \beta ^i}{dt}G_{ij}\beta^j >> \ne 0
\label{fluct}
\ee
in general,
which affects
energy uncertainty relations for the string.
\pr
(vi) The above picture leads naturally \cite{emnqm,emndollar}
to collapse of part of
the wave function describing observable modes,
in a way similar to that
proposed in ref. \cite{emohn} within a version
of the wormhole calculus, namely the asymptotic vanishing
(as $t \rightarrow \infty $) of the off-diagonal elements in space
of the matter density matrix $\rho $ :
\be
  \rho_{out} (x-x') = \rho _{in}(x,x') e^{-Dt(x-x')^2 + \dots}
\label{collapse}
\ee
where $x, x'$ are target spatial coordinates, and $D$ is proportional
to the sum of the squares of the
anomalous dimensions of the relevant perturbations of the
$\sigma$-model \cite{emndollar}.

\section{The Two-Dimensional String Black Hole Model}
\pr
As an illustration of our approach to non-critical string theory,
we now discuss the two-dimensional black hole model of ref. \cite{witt}.
We
regard it as a toy laboratory that gives us insight into the nature
of time in string theory and contributes to the physical effects
mentioned in the previous section.
\pr
The action of the model is
\be
   S_0=\frac{k}{2\pi} \int d^2z [\partial r {\overline \partial } r
- tanh^2 r \partial t {\overline \partial } t] + \frac{1}{8\pi}
\int d^2 z R^{(2)} \Phi (r)
\label{action}
\ee
where $r$ is a space-like coordinate and $t$ is time-like,
$R^{(2)}$ is the scalar curvature, and $\Phi$ is the dilaton field. The
customary interpretation of (\ref{action}) is as a string model with
$c$ = 1 matter, represented by the $t$ field, interacting with a
Liouville mode, represented by the $r$ field, which has $c  < 1$ and
is correspondingly space-like \cite{aben3,aben}.
As an illustration of the
approach outlined in the previous section, however, we re-interpret
(\ref{action}) as a fixed point of the renormalization group
flow in the local scale variable $t$. In our interpretation, the
``matter'' sector is defined by the spatial coordinate $r$, and has
central charge $c_m$ = 25 when $k  = 9/4$
\cite{witt}.Thus the model
(\ref{action}) describes a critical string in a dilaton/graviton
background. The fact that this is static, i.e. independent of $t$,
reflects the fact that one is at a fixed point of the renormalization
group flow \cite{emndollar,emnnew}.
\pr
We now outline how one can use the machinery
of the renormalization group in curved space,with $t$ introduced as
a local renormalization
scale on the world sheet, to derive the model (\ref{action}). A detailed
technical description is given in \cite{emnnew,emndollar}
There are two contributions
to the kinetic term for $t$ in our approach, one associated with
the Jacobian of the path integration over the world-sheet metrics, and
the other with fluctuations in the background metric.
\pr
To exhibit the former, we first choose the conformal gauge
$\gamma _{\alpha \beta}=e^{\rho}
{\hat \gamma}_{\alpha\beta}$ \cite{DDK,mm},
where $\rho$ represents the Liouville mode. We will later identify $\rho$
with an appropriate function of $\phi$, thereby making the local scale
$\phi$ a dynamical $\sigma$-model field. Ref.\cite{mm}
contains an
explicit computation of the Jacobian using heat-kernel regularization,
which yields
\be
-\frac{1}{48\pi}[\frac{1}{2}
\partial_\alpha \rho \partial^\alpha \rho +
R^{(2)}\rho + \frac{\mu}{\epsilon} e^\rho +
S'_G ]
\label{liouvillekin}
\ee
where the counterterms $S'_G$ are needed to remove the non-logarithmic
divergences associated with the induced world-sheet
cosmological constant term $\frac{\mu}{\epsilon}e^\rho$,
and depend on the background fields. This procedure
reproduces the critical string results of ref. \cite{witt}
when one identifies
the Liouville field $\rho$ with 2$\alpha'  \phi$. Equation
(\ref{liouvillekin})
contains a negative (time-like) contribution to the kinetic term
for the Liouville (time) field, but this is not the only such
contribution, as we now show.
\pr
We recall that the renormalization of composite operators in
$\sigma$-models formulated on curved world sheets is achieved by
allowing an arbitrary dependence of the couplings $g^i$ on the
world-sheet variables $z,{\bar z}$ \cite{shore,osborn}.
This induces counterterms of ``tachyonic'' form,
which take the following form in dimensional regularization with
$d  = 2 -  \epsilon$ \cite{osborn}:
\be
\int d^2z \Lambda_0
\label{tachloc}
\ee
where
\be
\Lambda _0=\mu^{-\epsilon}(Z(g)\Lambda + Y(g))
\label{ctr}
\ee
Here
$Z(g)$ is a common
wave function renormalization that maps target scalars into scalars,
$\Lambda$ is a residual
renormalization factor,
and the remaining counterterms $Y(g)$ can be expanded as power series
in 1/$\epsilon$, with the the one-loop result giving a simple pole.
Simple power-counting yields the following form for $Y(g)$:
\be
Y(g)=\partial _\alpha g^i {\cal G}_{ij} \partial ^\alpha g^j
\label{zamolmetric}
\ee
where ${\cal G}_{ij}$
is the analogue of the Zamolodchikov metric
\cite{zam} in this formalism, which is
positive for unitary
theories. It is related to the divergent part of the
two-point function $<V_i V_j>$ \cite{osborn} that
cannot be absorbed
in the conventional renormalization of the operators $V_i$. We need
to consider a $\sigma$-model propagating in a graviton background
$G_{MN}$, in which case a standard one-loop computation \cite{osborn}
yields the
following result for the simple $\epsilon$-pole in $Y$:
\be
Y^{(1)}=\frac{\lambda}{16\pi \epsilon}\partial _\alpha G_{MN}
\partial^\alpha G^{MN}
\label{polemetric}
\ee
where $\lambda\equiv 4\pi \alpha '$ is a loop-counting parameter.
We note that the wave-function renormalization
$Z(g)$ vanishes at one-loop.
In ref. \cite{osborn}
$G_{MN}$ was allowed to depend arbitrarily on the world-sheet
variables, and all world-sheet derivatives of the couplings were
set to zero at the end of the calculation. In our Liouville mode
interpretation, we assume that such dependence occurs only through
the local scale
$\mu (z,{\overline z})$,
so that
\be
\partial _\alpha g^i ={\hat \beta^i}\partial _\alpha \phi (z,{\bar z})
\label{part}
\ee
where ${\hat \beta }^i=\epsilon g^i + \beta ^i(g)$
and $\phi=ln \mu (z,{\bar z})$.
Taking
the $\epsilon \rightarrow 0$ limit, and
separating the finite and $O(\frac{1}{\epsilon})$ terms,
we obtain for the former
\be
O(1)-terms : \qquad
Res Y^{(1)}=\alpha '^2 R \partial_\alpha \phi
\partial^\alpha \phi
\label{scalarcurv}
\ee
where $R$ is the scalar curvature in target space, and we have used
the fact that the one-loop graviton $\beta$-function
is
\be
\beta _{MN}^G=\frac{\lambda}{2\pi}R_{MN}
\label{gravbeta}
\ee
The non-logarithmic divergent terms
\be
      \frac{1}{\epsilon} \beta^i {\cal G}^{(1)}_{ij} \beta^j
\label{quadratic}
\ee
do not contribute to the renormalization group, and can be
removed explicitly by target-space metric counterterms
\be
S_G =\frac{1}{\epsilon} G_{\phi \phi} \partial _\alpha \phi
\partial^\alpha \phi
+ \delta S(\phi, r)
\label{metric}
\ee
where the coefficients $G_{\phi\phi}$ are
fixed by the requirement
of cancelling the $\frac{1}{\epsilon}$ terms (\ref{quadratic}).
The $\delta S$ denotes arbitrary finite counterterms, which
are invariant under the simultaneous
conformal
rescalings of the fiducial world-sheet metric, ${\hat \gamma}
\rightarrow
e^{\sigma}{\hat \gamma}$, and local
shifts of the scale $\phi \rightarrow
\phi -\sigma$. This last requirement arises as in
the conventional approach to Liouville gravity \cite{DDK,mm}, where
the local renormalization scale $\phi$ is identified with the
Liouville mode $\rho$, after appropriate normalization. In our
interpretation
one is forced to treat the scale $\phi$ simultaneously as
the target time coordinate.
\pr
In the case of the Minkowski black hole model of ref. \cite{witt},
the
Lorentzian curvature is
\be
R=\frac{4}{cosh^2r}=4-4tanh^2r,
\label{curba}
\ee
which we substitute into equation (\ref{scalarcurv})
to obtain the form of the
second contribution to the kinetic term for the Liouville field $\phi$.
Combining the world-sheet metric Jacobian term in (\ref{liouvillekin})
with the
background fluctuation term
(\ref{scalarcurv},\ref{curba}),
we finally obtain the
following terms in the effective action
\be
\frac{1}{4\pi \alpha '} \int d^2z [
\partial _\alpha r \partial ^\alpha r -
tanh^2r \partial_\alpha \phi
\partial^\alpha \phi + dilaton-terms ]
\label{effectact}
\ee
Thus we recover the critical string $\sigma$-model action
(\ref{action})
for the Minkowski black-hole.
Dilaton counterterms are incorporated
in a similar way, yielding the dilaton
background of \cite{witt}. In addition,
as standard in stringy
$\sigma$-models, one also obtains the necessary
counterterms that guarantee target-space diffeomorphism
invariance of the Weyl-anomaly cefficients \cite{shore}.
Details are given in ref. \cite{emnnew}.
\pr
It should be noticed that
the renormalization group yields automatically the Minkowski
signature, due to the $c_m =25$ value of the matter central charge
\cite{aben3,aben}.
However,
as we
remarked in ref. \cite{emnnew,emndollar},
one can also switch over
to the Euclidean black hole model, and still maintain the
identification of the compact time with some appropriate function
of the Liouville scale $\phi$  that takes into account the
compactness of $t$ in that case. This Euclidean version is better
studied from the point of view of discussing exactly marginal
deformations that turn on matter in the model (\ref{action}).
In ref. \cite{chaudh} it was argued that the exactly marginal
deformation that turned on a static tachyon background for the
black hole of ref. \cite{witt} necessarily involved the
higher-level topological string
modes, which are non-propagating
delocalized
states.
This is a consequence
of the operator product expansion of the tachyon zero-mode operator
${\cal F} _{-\frac{1}{2},0} ^c    $ \cite{chaudh}:
\be
    {\cal F} _{-\frac{1}{2},0}^c    \circ
    {\cal F} _{-\frac{1}{2},0}^c    = {\cal F}_{-\frac{1}{2},0}^c
+ W_{-1,0}^{hw} + W_{-1,0}^{lw} + \dots
\label{ope}
\ee
where we only exhibit
the
appropriate holomorphic part for reasons of economy of
space.
The
$W$ operators and the $\dots$ denote level-one and higher string
states. The latter cannot be detected in local scattering
experiments, due to their delocalized character.
{}From a formal field-theoretic point of view such states cannot
exist as asymptotic states to define scattering, and also cannot
be integrated out in a local path-integral. They can only
exist as marginal deformations in a string theory.
An `experimentalist' therefore sees necessarily a
truncated matter theory, where the only deformation  is the
tachyon ${\cal F}_{-\frac{1}{2},0}^c    $, which is a (1,1) operator
in the black hole $\sigma$-model (\ref{action}), but is not
exactly marginal. This truncated theory
is non-critical, and hence
Liouville dressing
in the sense of (\ref{dressed}) is essential, thereby implying
time-dependence of the matter background.
Due to the
fact that the
appropriate
 exactly-marginal deformation associated with the tachyon in
these models
 involves all higher-level
string states, one can
conclude that in this picture the ensuing
non-equilibrium time-dependent backgrounds
are a consequence
of information carried off
by the unobserved topological string modes.
The r\^ole of the space-time singularity\footnote{We would like to
stress that the notion of `singularity' is clearly a low-energy
effective-theory concept. The existence of infinite-dimensional
stringy symmetries associated with higher-level string states
($W_\infty$-symmetries \cite{emn1}) `smooth out' the singularity,
and render the full string theory finite.}
 was crucial for this argument.
Indeed, in flat target-space matrix models \cite{matrix} the
tachyon zero-mode operator ${\cal F}$ is exactly marginal. As we shall
argue later on,   these flat models can be regarded as
an asymptotic ultraviolet
limit in time
of the Wess-Zumino black hole. Hence,
any time-dependence of the matter disappears in the vacuum,
leading to equilibrium.
\pr
In ref. \cite{emndollar} an additional deformation was considered
in parallel with the tachyon. This was the instanton vertex \cite{yung}
\be
   V_{I}             =g^I \int d^2 \rho \frac{1}{\rho ^4 } d^2z
e^{\rho\partial (sinhr e^{-it}) + hc}
\label{inst}
\ee
where $g^I << 1$ in
a dilute-gas approximation. The physics behind such a deformation
lies in the world-sheet interpretation of the Euclidean (Minkowski)
black hole
as representing a world-sheet vortex (spike) \cite{emnhall}. The charge
of such objects is proportional to the black-hole mass, and hence
the instantons, that induce transitions between defects of different
charge, can
represent mass changes
of black holes.
Higher-genus (quantum) effects are known \cite{emndec}
to produce black hole decay, and it is sensible to consider
instanton effects as being related to these. In addition, from
a conformal field theory point of view, it is known that deformations
consisting purely of
higher-level string modes do induce global rescalings to the black-hole
metric \cite{chaudh}, and hence
shifts in its mass. Therefore we associate
instanton deformations with local renormalization counterterms
representing collectively higher-level and higher-genus, i.e.
quantum string, effects.
\pr
Instantons in these models appear to have
logarithmically-divergent world-sheet actions and topological charge,
but their contribution to the path
integral is finite. From a renormalization point of view they are
by themselves
irrelevant operators. As we mentioned before,
their effect is similar  to that of higher
genera \cite{cohen}. They
induce \cite{emndollar} world-sheet
ultraviolet-divergent
(logarithmic)
shifts to the dilaton field,
and when considered
in the model deformed by a
tachyon, their effect is to lead to extra
infinities that  cannot be absorbed in conventional renormalization
of the tachyon fields. The presence of matter
is essential in this framework, given that
the instantons alone cannot affect the renormalization
group flow induced by relevant operators (tachyons).
Instead, in the matter-instanton deformed theory
one has a renormalization of the
Wess-Zumino level parameter, which, thus becomes scale-dependent.
It can be shown \cite{emndollar,emnnew} that
close to the ultraviolet  fixed point the renormalized $k_R$ in the
dilute-gas approximation is given by
\be
       k_R(t) \simeq e^{-4\pi \beta ^I T_0 t}
\label{kr}
\ee
where $t$ is the target time, and $T_0$ denotes the coupling  of the
(relevant) tachyon perturbation.
The
instanton $\beta$-function itself depends
on $k$, e.g. \cite{yung}
$\beta ^I =-\frac{k}{2}g^I$
in the case of
large $k$,
where the instantons form a dilute gas, giving
an exponential increase of $k_R$  with $t$ \cite{emndollar,emnnew}.
\pr
\section{Time-Dependent ``Fundamental Constants''}
\pr
The consequences of the
renormalization effect (\ref{kr}) of the effective
level parameter $k_R (t)$
are dramatic
for the physics of the string.
First, we recall that the
mass $M$
of the black hole is given in units of the
Planck mass
by \cite{witt,emndollar}
\be
   M/M_{planck}=\sqrt{\frac{1}{k_R(t)-2}}e^{const}
\label{mass}
\ee
implying asymptotically a vanishing-mass black-hole, i.e. flat
target-space time \cite{witt,emn1}. This makes the
connection with the
flat $c=1$ matrix model at the ultraviolet fixed point of
the flow.
\pr
Secondly, we note that the
exact target-space background
metric of the Wess-Zumino
$\sigma$-model (for finite $k$)
has the following asymptotic form for large $r \rightarrow \infty$
\cite{verlinde}
\be
   ds^2=2(k-2) (dr^2 - \frac{k}{k-2}dt^2 )
\label{effc}
\ee
In view of (\ref{kr}), the above relation implies
 an effective
time dependence of the velocity of light:
\be
                 c_q=c\sqrt{\frac{k(t)}{k(t)-2}}
\label{speed}
\ee
where $c \equiv c(\infty)$.
\pr
The result (\ref{speed}), implying an increase of the velocity
of light as one approaches the infrared fixed point,
can be re-derived from a local-renormalization-group
view-point. Consider the two-dimensional string model
(\ref{action})
perturbed by an irrelevant
deformation $g^i$. As shown in ref. \cite{emndollar}
and reviewed in section 3,
following the local-renormalization-scale formalism of
ref. \cite{shore,osborn} in dimensional regularization
$d=2 - \epsilon$,
an induced metric counterterm
for the time-component of the metric tensor assumes the form,
\be
        g^i {\cal G}^{(1)}_{ij}\beta^j \partial \phi {\overline
\partial }\phi
\label{counterterm}
\ee
where $\phi$ is the (time-like) Liouville field, and
${\cal G}_{ij}^{(1)}$ is the residue of the first pole
in $\epsilon$ of the Zamolodchikov metric in coupling constant
space. This term contributes to the wave function
renormalization of $\phi$.
For irrelevant deformations,
like instantons in the black-hole,
the $\beta ^i < 0$ and the  counterterm
(\ref{counterterm}) is negative. Taking into account
the discussion in section 3 showing that
in the case of two-dimensional black-hole backgrounds
the Liouville mode induces a kinetic term for $\phi$  of the form
$-tanh^2r \partial \phi {\overline \partial}\phi $ \cite{emnnew},
we observe
that (\ref{counterterm})
corresponds
to an additive shift in the time-component of the background metric.
Hence, in the limit $r \rightarrow \infty$ this would imply
an increase in the velocity of light.
The presence of a relevant operator, on the
other hand, leads to a decrease of the effective velocity of light.
In the Wilsonian procedure of renormalization, one integrates out
the irrelevant deformations, in the sense that one sets
the respective
$\beta$-functions to zero and solves the irrelevant couplings
in terms of the
relevant ones in order to
obtain
the known Gell-Mann-Low $\beta$-functions for the latter.
In the $\sigma$-model
approach, this would correspond to considering the tachyon perturbations
in the presence of an instanton background.
\pr
We now remark that
in the background defined by (\ref{effc}) one
may consider (low-energy) particle excitations
and derive the conventional Lorentz transformations
for coordinates
$r$ and time $t$
and particle velocities $u <c_q$.
The arrow of time does not appear in these Lorentz
transformations:
to see the arrow of time we must consider
transitions among theories characterized by different $k_R$,
in which case one abandons flat space-times and equilibrium situations.
Such transitions are induced by the interaction of the
propagating light string modes with the {\it `ether'} of the
topological string states.
Then Lorentz transformations
are only approximate, and should be
replaced by general coordinate transformations appropriate
for non-critical string theory. Such transformations
are viewed as local redefinitions of $\sigma$-model fields,
and undergo renormalization group flow.
The limit $c_q \rightarrow \infty$
appears `Galilean', but it
it should be remembered that in this limit the space-time
picture changes completely, since the
back-reaction effects
are enormous as one approaches the topological phase. The
dominance of delocalized
topological modes with discrete energies and
momenta in such a case is consistent with the
{\it instantaneous influence}
implied by
$c_q \rightarrow \infty$.
However, the above
perturbative framework, is certainly
inapplicable near the  topological phase. Thus, until
we understand the theory in the infrared,
the above remarks should be considered as
speculative.
\pr
The  result (\ref{speed})
is reminiscent of the well-known variation
of the velocity of light in a medium with its density and
temperature \cite{tarach}. Indeed, there the effective
velocity of light decreases as the temperature of the heat-bath
increases.
However, the mechanism here is different,
since the `medium' of non-local topological string modes does
not yield a plasmon mass.
\pr
The time-dependence of the string as it approaches the ultraviolet
fixed point is reflected in a computation of the string
position-momentum uncertainty relation in the $\sigma$-model
deformed by tachyons and instantons.
The result for the position-momentum uncertainty, defined appropriately
to incorporate curved gravitational backgrounds \cite{emnnew},
can be expressed as
\be
(\Delta X \Delta P)_{min} \equiv \hbar _{eff}(t) =\hbar (1 + O(\frac{1}{k(t)}))
\label{uncert}
\ee
where $\Delta A \equiv (<A^2> - <A>^2)^{\frac{1}{2}}$,
$< \dots >$ denotes $\sigma$-model vacuum expectation values, and
$\hbar$ is the critical-string Planck's constant.
The string uncertainty relation introduces a minimum
length $\lambda _s$ \cite{venez}, that
in our case decreases with time too :
\be
\lambda _s (t) \equiv (2\frac{\hbar (t) \alpha '(t)}
{c_q (t)^2})^{\frac{1}{2}}= \lambda _s^0
(1 + O(\frac{1}{k(t)}))
\label{length}
\ee
where the superscript $0$ a denotes quantity evaluated at the
ultraviolet fixed point in the
critical string theory. We also mention the following
important relation that can be proven \cite{emnnew}
while deriving (\ref{length}):
\be
\frac{\alpha ' (t)}{\alpha _0'}=\frac{c_q^2(t)}{c^2}
\label{alphacorr}
\ee
which stems from the relation between $k$ and
the Regge slope in the model (\ref{action}) \cite{emndollar,emnnew}.
The relation (\ref{alphacorr})
is consistent
with the
topological phase
transition that we conjecture occurs
at the infrared fixed-point of the flow, where $\alpha '(t)
\rightarrow \infty$ \cite{emntop}.

 \section{Applications to
 Inflation and the Cosmological ``Constant'' problem}
\pr
We saw
in the previous section  that fundamental constants
vary in the non-critical string Universe, because
of the interaction of the low-energy world with the topological
string modes. Fundamental ``constants'' attain fixed values
asymptotically, in the particular string vacuum which is
attained as the ultraviolet fixed point of the $\sigma$-model
renormalization group flow. So what physical meaning
is attached to the concept of the ``running'' velocity
of light $c_q (t)$
(\ref{speed}) or of the Planck ``constant''
((\ref{uncert}),(\ref{length})) ?
\pr
The velocity of light
can always
be absorbed in a rescaling of the time coordinate, provided
that it depends only on time and
that
this time dependence does not imply causality violations.
For instance, in the case of the asymptotic
black-hole metric (\ref{effc}),
one can redefine time $t \rightarrow t'(t)$
in such a way that
\be
  dt'= \sqrt{\frac{k}{k-2}}dt
\label{newtime}
\ee
This rescaling implies that
such a time dependence of the velocity of light
will be unobservable locally,
in the sense that in a co-moving frame the velocity
of light will appear as constant that can always be set to unity.
{}From a renormalization-group point of view, the redefinition
(\ref{newtime}) represents a local scheme change. As long as
there is a finite maximum speed in the low-energy world,
this will not have physical consequences locally.
However, we conjecture that the velocity of light
becomes ill-defined at the topological phase transition in the
infrared limit, where the normal concepts of space and time
break down.
\pr
However, the evolution of the velocity of light
may have important cosmological consequences.
If we consider a Friedmann-Robertson-Walker Universe, the
horizon distance $d$
in  co-moving coordinates over which an
observer can look back is
\be
    d=\int dt c_q (t) =\int dt \sqrt{\frac{k(t)}{k(t)-2}}
\label{horizon}
\ee
which is larger than the naive estimate $d=ct$, because the
effective $k(t)$ was smaller at earlier times. Indeed,
the horizon distance could even become infinite
if $k(t) \rightarrow 2$ in a suitable way as $t \rightarrow 0$, but
this conjecture takes us beyond the dilute-gas approximation
where we can compute reliably. Moreover, the time-dependence
(\ref{length}) of the minimum string length indicates
that the unit of length
in the low-energy
world appears to shrink with time. Both these features
are reminiscent of inflationary cosmology, although
we do not have an additional inflaton field.
\pr
The correspondence to inflation is reinforced by the
possible appearance of Jeans-like instabilities \cite{sv},
which may affect low-energy string modes at finite $k$, as we now argue.
We start from the observation that the $SL(2,R)/U(1)$ quantum
$\sigma$-model
metric (\ref{effc}) contains an overall scale factor
\be
a(t)^2= 2(k(t)-2)
\label{scalefactor}
\ee
which grows with time $ t $. This scale factor supplements
any time-dependent dilaton scale factor (cf the first, simpler, linear
example in ref. \cite{aben3})
that could be absorbed in the
metric background after appropriate redefinition.
For large times, and hence
large $k \rightarrow \infty$,
the background (\ref{effc})
is identical after the time redefinition $ t \rightarrow
e^{\delta t }$, $\delta \equiv -4\pi \beta ^I T_0$,
to the linearly-expanding universe
of ref. \cite{sv}, for which there are no
Jeans-like instabilities.
However,
this is no longer the case if $\beta ^I T_0$ is not constant and
$k_R(t)$ is no longer a simple exponential, as could occur away
from the ultraviolet fixed point when instanton effects
become strong. As is shown elsewhere \cite{emnelse}, for general
$k_R (t)$, the evolution equation for $a(t)$ in the cosmic time frame
is
\be
\frac{\ddot a}{a}=\frac{1}{2k_R    (k_R-2) }\{ \partial_t ^2
k_R
-\frac{k_R-1}{k_R (k_R-2)}( \partial _t  k_R   )^2 \}
\label{general}
\ee
where the dot denotes differentiation with respect to
the cosmic time $t_c$, defined as
\be
dt_c=a(t')dt'
\label{proper}
\ee
The first factor in the square brackets in (\ref{general}) might
dominate the second near the infrared fixed point, in which case
$\frac{\ddot a}{a} > 0$ and a Jeans-like instability occurs.
\pr
We re-emphasize, though, that a physical expansion of the Universe
occurs whether or not Jeans instabilities appear. It is natural
to enquire into their physical meaning, should they appear.
To understand this, it is better
to go to higher dimensions, where
there are transverse modes. It can be shown \cite{sv} that
the temporal evolution of these modes is similar to the longitudinal
ones.
The point is that due to the existence of these unstable modes,
the proper size of the string
does not grow in the same way as the expansion of the Universe.
Hence strings stretch as one goes back in time
\footnote{There is
no stretching in two space-time dimensions. However, even in
that case the existence of unstable
modes in the longitudinal amplitude,
makes things similar to the
higher-dimensional case.}. We expect that at early
times, close to the topological string phase \cite{emntop},
there will be string regeneration
through the breaking of strings larger than the
Hubble horizon size $R_0$ \cite{turok}. In
our picture, these effects would be
attributed to purely quantum effects of the string, given the
connection of the instanton perturbations with higher-genus
instabilities of the black  hole background \cite{emndec}.
This
quantum string
regeneration
scenario is similar to the string-driven inflation
proposed by Turok \cite{turok}.
Indeed,
were it not for
this regeneration phenomenon, the exponential expansion of the
Universe in an inflationary scenario would
imply
a strong suppression
of the string density, due to conformal stretching of the string size
which grows
like the scale factor of the Universe.
\pr
In this simplified scenario, one  can have a qualitative
picture of the entropy production rate in the Universe.
In our framework,
the rate of entropy increase with
time
is given by \cite{emnqm,emnhall}
\be
\partial _t S = \beta ^i G_{ij} \beta ^j S \qquad ; \qquad
G_{ij}=<V_i V_j >
\label{entropy}
\ee
where the unitarity requirement of the world-sheet theory implies
the positivity of the Zamolodchikov metric \cite{zam} $G_{ij} >0$.
Using the C-theorem \cite{zam}, especially in
its string formulation \cite{mavc} on the fiducial-metric world-sheet,
one may write
\be
 \beta ^i G_{ij} \beta ^j =-\frac{1}{12}\partial _t C(g) \qquad :
\qquad
 C(g) =   \int d^Dy \sqrt{G} e^{-2\Phi} <TT> + \dots
\label{ctheorem}
\ee
 In
this expression, the
$y$ denote  target spatial coordinates,
$\Phi$ is the dilaton field, and $T \equiv
T_{zz}$ is a component of the world-sheet stress tensor.
The $\dots $ denote the remaining two-point functions that appear in the
Zamolodchikov C-function \cite{zam}, which
involve the trace $\Theta $
of the stress tensor, i.e.
$<T \Theta >$ and $ <\Theta \Theta >$. Taking into account the
off-shell corollary of the C-theorem, $\frac{\delta C(g)}{\delta g^i}
=G_{ij}\beta ^j$, it can readily be shown \cite{mavr} that
such terms
can always be removed by an appropriate renormalization-scheme choice,
that is by
appropriate redefinitions of the
renormalized couplings $g^i$, and hence play no r\^ole in the physics.
Thus, one can
solve (\ref{entropy}) for the entropy $S$
in terms of the Zamolodchikov
C-function
\be
S(t)=S_0 e^{-\frac{1}{12}\int _0^t \int d^Dy \sqrt{G}e^{-2\Phi}
<TT> + \dots}
\label{entrzam}
\ee
where the minus sign in the exponent indicates the opposite
flow of the time $t$ with respect to the renormalization-group
flow.
Expression
(\ref{entrzam}) reduces a complicated target-space computation
of entropy production in an inflationary scenario to a
conformal-field-theory computation of two-point functions
involving components of the stress tensor of a first-quantized string.
We observe from (\ref{entropy}) that
the rate of entropy increase
is maximized on the
maximum-$\beta ^i$
surface in coupling constant
space.
At late stages of the inflationary era,
i.e. close to the ultraviolet fixed point, the rate of change of $S$
is strongly suppressed, due to the smallness of the
$\beta ^i$.
\pr
The non-critical string  scenario for the expanding
Universe described in the preceeding paragraphs
offers the prospect of solving
the three basic
problems of the standard-model cosmology
 in a manner reminiscent
of conventional inflation \cite{guth}.
The {\it horizon problem} could be solved by the
enhanced look-back distance (\ref{horizon}), and/or
the breakdown of the normal concepts of space and
time in a transition to a topological phase close to
the infrared fixed point. The {\it flatness}
problem could be solved by an epoch of
exponential expansion, induced by a Jeans-like instability
in (\ref{general}). The {\it entropy problem} could be solved by
the enhanced rate of entropy production (\ref{entrzam})
at early times.
However,
the
crucial difference in our
approach is that the fundamental scalar field, usually
termed the
{\it inflaton},
is replaced by
a world-sheet field, the Liouville mode,
in our approach.
Fluctuations of this field create the renormalization
group flow of the system that leads to the generation
of propagating
matter, in the way described above and
in previous works\cite{emnhall,emndollar}.
Of course, this mode is associated with the appearance of
a target space scalar, the dilaton, but the latter is part of the
metric background. This can be seen
clearly in the two-dimensional
Wess-Zumino string theory, which may be considered as a
prototype for the description of a spherically-symmetric
($s$-wave) four-dimensional
Universe \cite{emn4d}. In this model
the dilaton belongs to the graviton level-one string multiplet,
which is a non-propagating (discrete)
string mode, and as such can only
exist as a background, in contrast to a massless `tachyon' mode,
which propagates and scatters.
\pr
As a final comment, we discuss
the
cosmological ``constant'' problem in our framework.
We adopt a toy-model approach,
and concentrate
on the one-loop $\beta$-functions for the graviton and dilaton
fields only. Tachyonic effects are
summarized by the  running
of the level parameter,
and we shall not deal with the explicit
form of the tachyon potential. As argued in \cite{banks},    its
particular form is renormalization-scheme dependent and, thus,
irrelevant for our purposes. We also remark that, as will
become clear from the analysis below, {\it our results can be
extended to any number of dimensions for the target space},
provided one maintains the requirement of
maximal space symmetry.
\pr
Consider the following
one-loop results for the dilaton and graviton
$\beta$-functions in bosonic $\sigma$-models \cite{tseytlin,osborn}:
\bea
   \beta ^\Phi \equiv \frac{d \Phi }{d \phi}
   &=& -\frac{2}{\alpha '} \frac{\delta c}{3} + \nabla ^2 \Phi
- (\nabla \Phi )^2  \nn \\
  \beta ^G _{MN} \equiv \frac{d G_{MN}}{d\phi}
  &=&- R_{MN} - 2 \nabla _M \nabla _N \Phi
\label{betafunceq}
\eea
where $\phi$ denotes our covariant Liouville cutoff (c.f. the relative
minus sign compared
with the notation
of ref. \cite{tseytlin,osborn} where the cutoff is defined
with the dimensions of mass), and
$\delta c = c-26$, the $26$ coming from the space-time
reparametrization
ghosts.
If
the central charge of the
theory is not 26, as is the case of non-critical bosonic strings,
then
a
cosmological constant term appears
in the target space effective action.
The form of this target-space
action,
whose variations
yield the $\beta$-functions
(\ref{betafunceq}), reads:
\be
 {\cal I}=\frac{2}{\alpha '}
 \int d^Dy \sqrt{G}e^{-\Phi} \{ \frac{1}{3}\delta
 c - \alpha '
(R + 4 (\nabla \Phi )^2 + \dots ) \}
\label{effbossigma}
\ee
where the $\dots$ denote other fields in the theory that we shall not
use explicitly. We now notice that the effects of the tachyons
in our two-dimensional target-space string model amount to a shift
of the level parameter $k(\phi)$ with the
renormalization
group scale. This is the result of the combined
effects of tachyon and instanton deformations, the latter
representing higher-genus instabilities. The instantons alone, as
irrelevant deformations,
produce an initial instability by inducing an
increase of
the central charge,
which then flows
downhill towards 26
in the presence of relevant matter (tachyon) couplings.
Hence there is
a running central charge $c(\phi) > 26$, according to
the C-theorem \cite{zam}, that will, in general, imply a non-vanishing,
time-varying (running), {\it positive}
cosmological constant, $\Lambda (\phi)$,
for the background of (\ref{effc}). Its precise
form is determined by consistency with the equations (\ref{betafunceq}).
\pr
For simplicity,
we assume that the only effect of the dilaton
is a constant contribution to the scale anomaly, which is certainly
the case of interest. This allows one to decouple
$\Phi$ in the
field equations obtained from (\ref{effbossigma}). Then
the latter read
\bea
\frac{\delta {\cal I}}{\delta \Phi}&=&\Lambda (\phi) - R  \nn \\
\frac{\delta {\cal I}}{\delta G_{MN}}&=&-R_{MN} + \frac{1}{2}G_{MN}R
\eea
In two dimensions the second equation is satisfied identically.
Decoupling of the dilaton field also implies that the first equation
yields
\be
            R=\Lambda (\phi)
\label{curvcosmo}
\ee
We note at this stage that the fact that the background
equation for graviton is satisfied automatically implies that the
metric in coupling constant space of the full two-dimensional
target-space string is singular. This seems to be in agreement
with the time-like signature of the Liouville field,
which implies
non-unitary partial contributions to the $\sigma$-model, and should
be compared with the situation of ref. \cite{aben3}, between
compact spaces and four dimensional uncompactified Minkowski
space. In the present case, the conformal field theory associated with
the space coordinate $r$ and the running coupling $G_{rr}(\phi)$
is unitary and satisfies the $C$-theorem
\cite{zam}. The non-unitarity of the Liouville part is essential
in assuring the
Weyl invariance of the full theory.
\pr
The metric background (\ref{effc}) has a maximal symmetry in its
space part. To make the analysis more general, we extend the
background to $d=2 + \epsilon$ dimensions, keeping the maximal
symmetry in the spatial part of the metric \footnote{The
$G_{00}$-component depends  at most on the time $\phi$ and can be
absorbed in a redefinition of the time variable. It will not
be of interest to us here.}:
\be
      G_{ij}=a^2(\phi){\hat G}_{ij}(x) \qquad ; \qquad
R_{ij}=\frac{R}{d-1}G_{ij}
\label{maximal}
\ee
where the scale factor $a^2(\phi)$ is related to $k(\phi)$ as in
(\ref{effc}). The curvature of $G_{ij}$ is assumed to be that
of a scaled sphere
\be
            R=\frac{b}{d-2}
            \frac{1}{a^2(\phi)}=
            \frac{b}{\epsilon}
            \frac{1}{a^2(\phi)}
\label{rhat}
\ee
where $b$ is a constant.
{}From (\ref{curvcosmo}) one finds the following
relation between the
cosmological constant and scale factor
\be
            a^2(\phi)=\frac{b/(d-2)}{\Lambda (\phi)}
\label{scalecosmo}
\ee
Substituting into the $\beta$-function equation (\ref{betafunceq})
for the spatial
component of $G_{ij}$, it is immediate to see
that the $d$-dependent factors cancel out, leaving an effective
equation for the running cosmological constant
\be
\frac{1}{\Lambda ^2}\frac{d \Lambda}{d \phi}  =\frac{1}{d-1}
\label{effcosmo}
\ee
which should be applicable
to higher dimensions.
As we have explained above,
the actual target-time flow is opposite to that of the
renormalization group, and hence
it is the negative real $\phi$-axis
that is relevant,
given also that $\phi =ln(a^2/L^2)$ , where $a$ and $L$ are
ultraviolet and infrared length cut-offs respectively.
This yields as a solution for $\Lambda (\phi)$\footnote{We remark
that
a similar
equation
has also been
considered in ref. \cite{kogan2}, but the flow  of time in that
reference coincides
with the renormalization group flow. In such a case,
one gets sensible results only for negative initial values of the
cosmological constant, contrary to our case where we have a
vanishing cosmological constant asymptotically, starting from
positive initial values.}
\be
    \Lambda (t)=\frac{\Lambda (0)}{1 + t
     \frac{\Lambda (0)}{d-1}} \qquad ; \qquad t\equiv -\phi > 0
\label{runnincosmo}
\ee
which for positive $\Lambda$ implies an
asymptotically-free
cosmological constant $\beta$-function, thereby leading to
a vanishing cosmological constant at the ultraviolet fixed point
on the world-sheet.
\pr
The rate of the decrease of $\Lambda (\phi)$
is determined by its initial value at the infrared
fixed point, where we conjectured that
the theory makes a transition to a topological (twisted
$N=2$ supersymmetric) $\sigma$-model. It is of great
interest to estimate
this value in our two-dimensional model. This
can be done by noticing that
\be
        \Lambda (0)=\frac{2}{\alpha '(0)}\frac{ c(0)-26}{3}
\label{irlam}
\ee
where $c(0)-26
 =\frac{3 k(0)}{k(0)-2} -27 \simeq \frac{3k(0)}{k(0) -2}$,
given that $k(\phi) \rightarrow 2 $ as $\phi \rightarrow 0$. Thus,
taking into account (\ref{alphacorr}) one observes that
$\Lambda (0)$ is determined by the critical string tension, as it
should be, given that $\alpha '_0$
is the only scale in the problem
(or equivalently the minimal  string length): the result is
\be
     \Lambda (0) = \frac{2}{\alpha '_0}
\label{critalpha}
\ee
The latter result implies a really fast decay of the cosmological
constant in this model. Notice that the finite initial
value of $\Lambda (0)$ implies from (\ref{curvcosmo})
 no curvature singularity in the
Euclidean model at the origin of target space $r=0$,
as is indeed the case of the two-dimensional
black-hole model of ref. \cite{witt}, given that
this point
is a pure coordinate singularity. The above analysis
for the cosmological constant, therefore, applies most likely
to singularity-free inflationary universes.
It is understood that until the
precise behaviour of the running couplings near the infrared fixed
point is found, there will always be uncertainties in the above
estimates. String perturbation theory is not applicable near the
topological phase transition, and the infinities we get in the various
running couplings constitute
an indication of this. In the complete theory,
these infinities should be absent.
\pr
At this stage,
we notice that such a
scenario for an asymptotic (in time)
vanishing of the cosmological constant was conjectured
in ref. \cite{aben3}, but no explicit example was provided.
In our case, it is the association of the Liouville mode
with a (local) renormalization group scale, flowing
irreversibly, that makes possible a realization of such a scenario
via (\ref{runnincosmo}).
\pr
In order to asses the physical consequences of the
equation (\ref{runnincosmo}) for the asymptotic
vanishing of the cosmological ``constant'', we also need
to explore further the relationship
between the renormalization group time $t=-\phi$ and the physical
time measured, for instance, by atomic clocks. As already
mentioned, a time-dependent scale factor appears (\ref{effc})
already in the $\sigma$-model metric, supplementing any possible
time-dependent dilaton scale factors which may appear
\cite{aben3} in the ``physical'' metric. The latter is defined
as a co-moving frame, where such a time-dependent scale factor
is absorbed in an appropriate redefinition (\ref{newtime},\ref{proper})
so as to yield a conventional Friedmann-Robertson-Walker form
for the target-space metric:
\be
ds^2=2(k(t'')-2)dr^2-(dt'')^2
\label{distono}
\ee
where $t''$ denotes the redefined time variable. In the previous
two-dimensional example, where the Wess-Zumino
model's dilaton field is space like outside
the horizon \cite{witt}, the redefined time $t''$ (\ref{distono})
is given simply by the scale factor (\ref{scalefactor})
\be
     dt'' =\sqrt{2 k(t   )} dt
\label{tdistono}
\ee
To integrate (\ref{tdistono}), we need to know the full
$t$-dependence of $k(t)$, which
is lacking away from the ultraviolet limit.
\pr
It is premature to discuss quantitatively the implications of the
vanishing mechanism (\ref{runnincosmo}), for this and other reasons.
However, we finish this talk by noting a few relevant points.
One is that the present age of the Universe in ``physical'' time is about
$10^{60}$ in natural (Planck) units. Another is that one-loop
calculations at the point-like field theory level in no-scale
supergravity models \cite{noscale,barrow})
yield a negative contribution
to
the cosmological constant that is $O((m_w/M_{Plnack})^4)  \simeq
10^{-60}$ in Planck units. Finally, we note that
astrophysical and cosmological observations are compatible
with a present-day value of the cosmological constant that is about
$10^{-120}$ in Planck units. Moreover, a cosmological constant
of this order of magnitude could even be a welcome adjunct
to Cold Dark Matter models. Thus it may even be desirable
that the cosmological ``constant'' has not yet completely relaxed.
\pr
\nk {\Large{\bf  Acknowledgements}}
\pr
The work of N.E.M. is supported by a EC Research Fellowship,
Proposal Nr. ERB4001GT922259 .
That of D.V.N. is partially supported by DOE grant
DE-FG05-91-GR-40633.

\end{document}